\documentclass[11pt,preprint2]{aastex}

\usepackage[]{natbib}
\usepackage{graphicx}
\usepackage{amssymb}
\usepackage{color}
\usepackage{ulem}

\newcommand{\etacar}{$\eta$~Car}

\input{macro.inp}

\slugcomment{astro-ph ver. 1.0}

\shorttitle{\SUZAKU\ Monitoring of $\eta$ Carinae}
\shortauthors{Hamaguchi et al.}

\begin{document}

\title{\SUZAKU\ Monitoring of Hard X-ray Emission from $\eta$ Carinae over a Single Binary Orbital Cycle}

\author{Kenji Hamaguchi\altaffilmark{1,2}, Michael F. Corcoran\altaffilmark{1,3},
Hiromitsu Takahashi\altaffilmark{4}, Takayuki Yuasa\altaffilmark{5}, Manabu Ishida\altaffilmark{6}, 
Theodore R. Gull\altaffilmark{7}, Julian M. Pittard\altaffilmark{8}, Christopher M. P. Russell\altaffilmark{9},
Thomas I. Madura\altaffilmark{7,10}}

\altaffiltext{1}{CRESST and X-ray Astrophysics Laboratory NASA/GSFC,
Greenbelt, MD 20771, USA}
\altaffiltext{2}{Department of Physics, University of Maryland, Baltimore County, 
1000 Hilltop Circle, Baltimore, MD 21250, USA}
\altaffiltext{3}{Universities Space Research Association, 7187 Columbia Gateway Drive, Columbia, MD 21046, USA}
\altaffiltext{4}{Department of Physical Sciences, Hiroshima University, Higashi-Hiroshima, Hiroshima 739-8526, Japan}
\altaffiltext{5}{Nishina Center, RIKEN, 2-1, Hirosawa, Wako, Saitama, Japan, 351-0198, Japan}
\altaffiltext{6}{Institute of Space and Astronautical Science/JAXA, 3-1-1 Yoshinodai, Sagamihara, Kanagawa 229-8510, Japan}
\altaffiltext{7}{Astrophysics Science Division, NASA Goddard Space Flight Center, Greenbelt, MD 20771, USA}
\altaffiltext{8}{School of Physics and Astronomy, The University of Leeds, Woodhouse Lane, Leeds LS2 9JT, UK}
\altaffiltext{9}{Bartol Research Institute, Department of Physics and Astronomy, University of Delaware, Newark, DE 19716, USA}
\altaffiltext{10}{NASA Postdoctoral Program Fellow}

\begin{abstract}
The \SUZAKU\ X-ray observatory monitored the supermassive binary system $\eta$~Carinae 10 times during
the whole 5.5~year orbital cycle between 2005$-$2011.
This series of observations presents the first long-term monitoring of this enigmatic system in the extremely 
hard X-ray band between 15$-$40~keV.
During most of the orbit, the 15$-$25 keV emission varied similarly to the 2$-$10~keV emission,
indicating an origin in the hard energy tail of the \KT\ $\sim$4~keV wind-wind collision (WWC) plasma.
However, the 15$-$25~keV emission declined only by a factor of 3 around periastron
when the 2$-$10~keV emission dropped by two orders of magnitude due probably to an eclipse of the WWC plasma.
The observed minimum in the 15$-$25 keV emission occurred after
the 2$-$10 keV flux had already recovered by a factor of $\sim$3.
This may mean that the WWC activity was strong, but hidden behind the thick primary stellar wind during the eclipse.
The 25$-$40~keV flux was rather constant through the orbital cycle, at the level measured with \INTEGRAL\ in 2004.
This result may suggest a connection of this flux component to the $\gamma$-ray source detected in this field.
The Helium-like Fe $K\alpha$ line complex at $\sim$6.7~keV became strongly distorted toward periastron as seen in the previous cycle.
The 5$-$9~keV spectra can be reproduced well with a two-component spectral model, which includes plasma in collision equilibrium (CE) and a plasma
in non-equilibrium ionization (NEI) with $\tau \sim$10$^{11}$~cm$^{-3}$~s$^{-1}$.
The NEI plasma increases in importance toward periastron.
\end{abstract}

\section{Introduction}

Most massive stars are found in binary systems \citep{Chini2012,Sana2012}.
When the companion is an early type star, the collision of their stellar winds (wind-wind collision: WWC) 
produces strong shocks and thermalizes gas to tens of millions of degrees Kelvin.
This hot gas emits X-rays, which are a good probe of the wind nature and interaction mechanism.
The shocks can also accelerate electrons to GeV energies, which produce radio synchrotron emission \citep[e.g.,][]{Pittard2006}.
These non-thermal electrons are also suspected to up-scatter UV emission from the stars through the inverse-Compton process to X-ray (and higher) energies.

Eta Carinae \citep[$d \sim$2.3~kpc,][]{Smith2006b} is a nearby example of an extremely massive binary system with energetic WWC activity \citep{Corcoran1997,Damineli1997,Ishibashi1999}.
The primary star is suspected to have had an initial mass of $\gtrsim$100~\UNITSOLARMASS\ \citep[see][]{Davidson1997,Hillier2001}
and is currently in the poorly understood Luminous Blue Variable (LBV) stage.
Since a series of eruptions between 1838$-$1890,
the two stars have been enshrouded by bipolar ejecta called the Homunculus Nebula (HN),
but their highly eccentric orbit ($e\sim$0.9) with a period of 5.54 years can be measured from periodic variations at various wavelengths \citep{Corcoran2005,Damineli2008}.
The companion star has not been detected directly, but it is believed to be an O supergiant or WN star \citep{VernerE2005a}.
The primary star has a thick slow wind with $v_{wind} \sim$420~\UNITVEL\ and \Mdot\ $\sim$8.5$\times$10$^{-4}$~\UNITSOLARMASSYEAR\ \citep{Groh2012},
while the secondary star has a thin fast wind with $v_{wind} \sim$3000~\UNITVEL\ and \Mdot\ $\sim$ 10$^{-5}$~\UNITSOLARMASSYEAR\ \citep{Pittard2002}.

The WWC of \etacar\ produces luminous X-ray emission from hot plasma up to \KT\ $\sim$4~keV, which has been observed mostly in the 2$-$10~keV band.
The emission increases inversely-proportional to the stellar separation, as suggested by WWC theory \citep{Stevens1992}.
However, the X-ray flux suddenly drops to a minimum level \citep{Corcoran2010} after reaching a maximum brightness.
Detailed studies \citep{Hamaguchi2007b,Hamaguchi2014a} revealed two distinct phases during the X-ray minimum ---
the deep X-ray minimum, which has the lowest observed flux level and lasts approximately three weeks,
and the shallow X-ray minimum, where the emission abruptly increases three-fold from the deep minimum level.
The deep minimum is probably produced by an eclipse of the WWC apex by the primary stellar body or wind, 
while the shallow minimum probably indicates the intrinsic decline of the WWC activity \citep{Hamaguchi2014a}.

\begin{deluxetable}{llllcccccc}
\tablecolumns{10}
\tablewidth{0pc}
\tabletypesize{\scriptsize}
\tablecaption{\SUZAKU\ Observation Log\label{tbl:obslogs}}
\tablehead{
\colhead{Abbr}&
\colhead{Obs~ID}&
\multicolumn{2}{c}{Time}&
\colhead{NP}&
\multicolumn{3}{c}{XIS}&
\multicolumn{2}{c}{HXD}\\  \cline{3-4}\cline{6-8}\cline{9-10}
\colhead{}&
\colhead{}&
\colhead{Date}&
\colhead{$\phi_{\rm X}$}&
\colhead{}&
\colhead{Exp}&
\colhead{Sensor}&
\colhead{SCI}&
\colhead{Exp}&
\colhead{Epoch}\\
\colhead{}&
\colhead{}&
\colhead{}&
\colhead{}&
\colhead{}&
\colhead{(ksec)}&
\colhead{}&
\colhead{}&
\colhead{(ksec)}&
\colhead{}
}
\startdata
SUZ050829&100012010&2005~08~29&1.389&XIS&49.8&0123&off&56.0&1\\
SUZ060203&100045010&2006~02~03&1.468&XIS&21.4&0123&off&18.1&1\\
SUZ070623&402039010&2007~06~23&1.717&HXD&54.7&013&on&51.6&3\\
SUZ080610&403035010&2008~06~10&1.891&HXD&35.5&013&on&27.2&4\\
SUZ081210&403036010&2008~12~10&1.982&HXD&48.5&013&on&42.4&5\\
SUZ090125&403037010&2009~01~25&2.005&HXD&28.8&013&on&17.5&5\\
SUZ090215&403038010&2009~02~15&2.015&HXD&35.6&013&on&31.1&5\\
SUZ090610&404038010&2009~06~10&2.072&HXD&51.2&013&on&49.1&5\\
SUZ091121&404039010&2009~11~21&2.153&HXD&49.4&0$^{\prime}$13&on&34.3&6\\
SUZ110724&406039010&2011~07~24&2.454&XIS&42.0&0$^{\prime}$13&on&49.1&11\\
\enddata

\tablecomments{
Abbr: Abbreviation adopted for each observation. 
Obs ID: Observation identification number of each observation.
Time: Observation start date and orbital phase. 
$\phi_{\rm X}$ = (observation start in Julian date $-$ 2450799.792)/2024 \citep{Corcoran2005}.
NP: Nominal Pointing position.
XIS/Exp: XIS exposure time.
XIS/Sensor: XIS sensors in operation. 0$^{\prime}$: One eighth of the XIS0 chip does not work.
XIS/SCI: Spaced Charge Injection operation.
HXD/Exp: HXD/PIN exposure time.
HXD/Epoch: Epoch of the HXD/PIN response file.
}

\end{deluxetable}

There have been several observations of \etacar\ in the hard X-ray band above 10~keV, up to $\sim$100~keV.
\citet{Viotti2002,Viotti2004} claimed a detection of extremely hard X-ray emission from \etacar\ with the PDS instrument on \SAX, 
but the measured flux was significantly higher than those of later measurements,
so source confusion in the wide PDS field of view ($\sim$1.3$^{\circ}$ FWHM) was suspected.
\citet{Leyder2008,Leyder2010} detected a flat power-law ($\Gamma \sim$1$-$2) source between $\sim$20$-$100~keV with \INTEGRAL/ISGRI.
They constrained the source position to within 1.6\ARCMIN\ of \etacar.
Since they found no X-ray source in a \CHANDRA\ image consistent with the observed spectrum above 20~keV,
they identified the source as \etacar.
\citet{Sekiguchi2009} analyzed the first two \SUZAKU\ \citep{Mitsuda2007} observations of \etacar\ around apastron in 2005
and detected X-ray emission between 15$-$40~keV with the HXD/PIN instrument.
They showed that the spectrum below $\sim$20~keV can be reproduced by \KT\ $\sim$4~keV plasma emission observed below $\sim$10~keV,
while the spectrum above 10~keV requires a flat power-law of $\Gamma \sim$1.4.
These papers suggested that the power-law component may originate from the inverse-Compton up-scattering of stellar UV photons 
by non-thermal GeV electrons accelerated at the WWC region.
On the other hand, the \AGILE\ and \FERMI\ $\gamma$-ray observatories discovered 
a relatively stable $\gamma-$ray source between 0.1$-$100~GeV \citep{Tavani2009,Abdo2010},
whose spectrum may be connected to this extremely hard X-ray source \citep{Farnier2011,Reitberger2012}.

The \SUZAKU\ observatory monitored \etacar\ 10 times between 2005$-$2011 and throughout one orbital cycle of \etacar.
\SUZAKU\ has the lowest background in the 15$-$40 keV band of any X-ray observatory launched before 2012, so that
it gives the most reliable results on the orbital modulation of extremely hard X-ray emission from \etacar.
It also has good sensitivity and spectral resolution between 5$-$9~keV, providing detailed profiles of
the $K\alpha$ and $K\beta$ line complexes of highly ionized Fe and Ni atoms.
In this paper, we present the flux and spectral variation of \etacar\ between 5$-$40 keV with orbital phase,
fit all the spectra with a consistent model, and discuss the nature of the observed emission components.

\begin{figure*}[t]
\begin{center}
\includegraphics[width=0.85\textwidth]{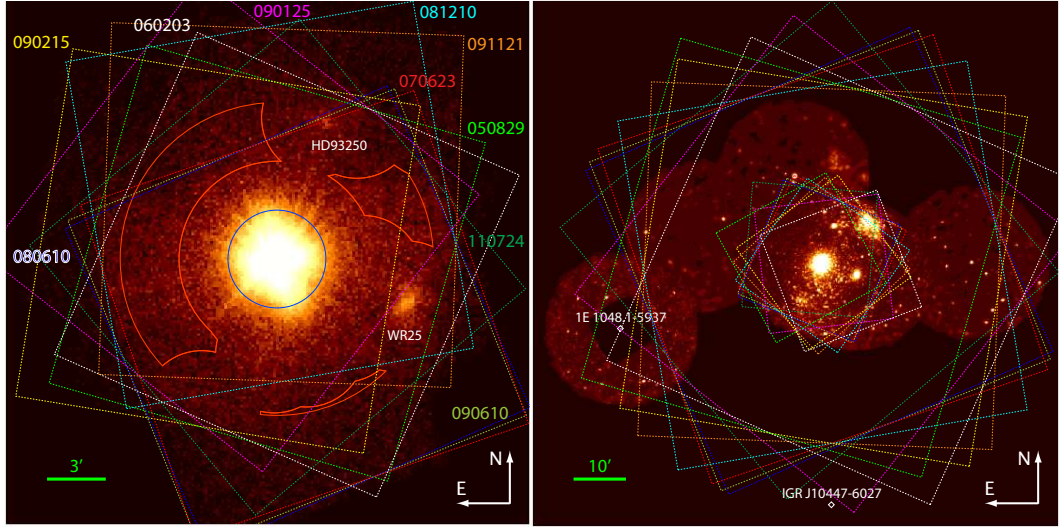}
\caption{{\it Left}: Composite XIS3 image of \etacar\ between 5$-$10~keV.
The solid blue circle at the center and the solid red regions are the source and background regions for the XIS analysis, respectively.
The dotted squares show the XIS {\FOV}s of individual observations.
{\it Right}: Approximate HXD/PIN {\FOV}s overlaid on a mosaic \XMM/MOS image of the Carina nebula between 2$-$7~keV.
The inner and outer boxes are boundaries of half and zero photons to the PIN detector from an on-axis source, respectively.
\label{fig:obsimg}
}
\end{center}
\end{figure*}

\section{Observations and Analysis}

\subsection{Observations}

\begin{figure}[h]
\plotone{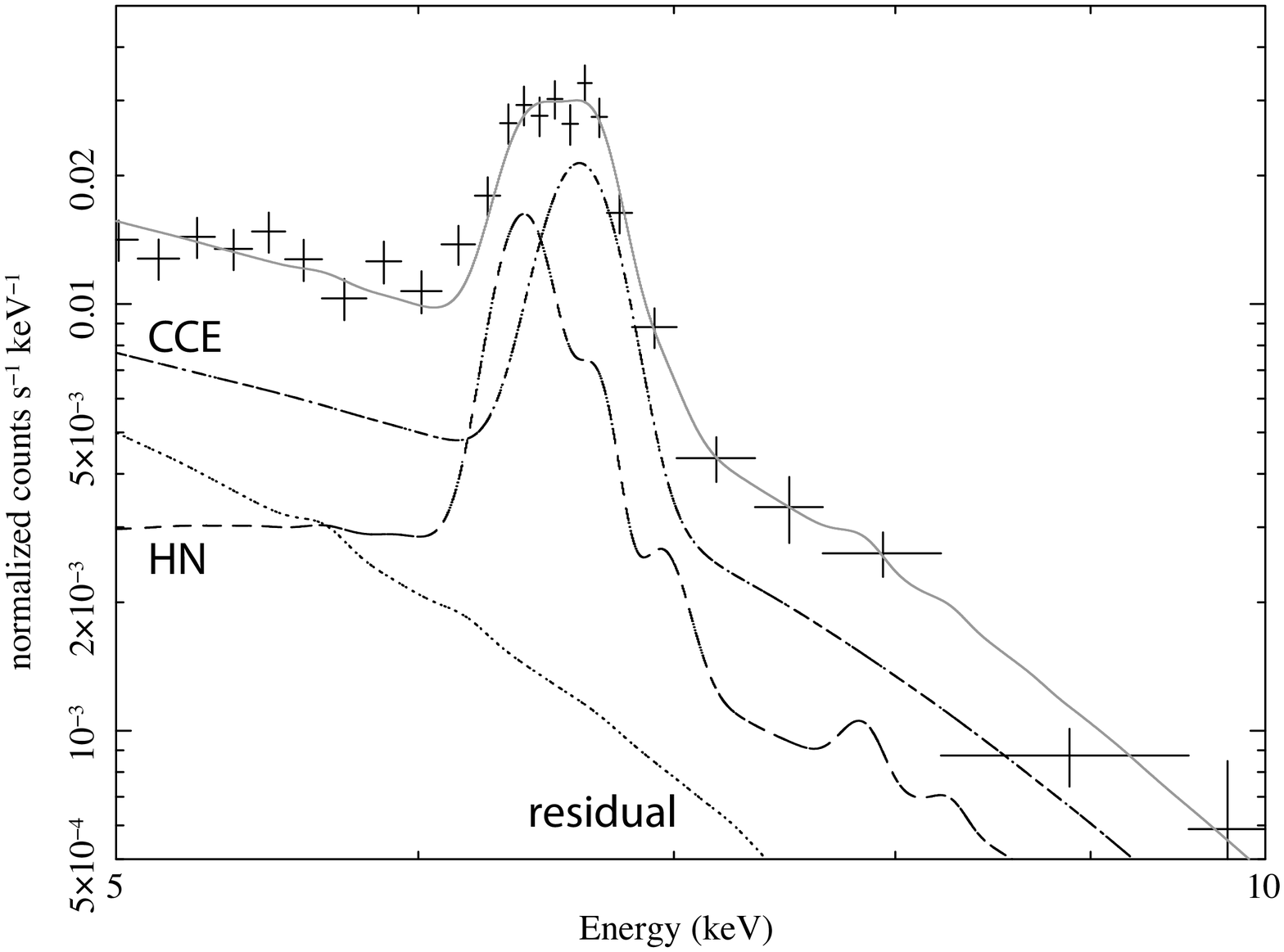}
\caption{
XIS0+3 spectrum of \etacar\ during the deep minimum (SUZ$_{090125}$).
The best-fit models of the \CHANDRA\ deep minimum spectra of the HN ({\it dash}, Hamaguchi et al. in prep.) and the CCE \citep[{\it dash-dot},][]{Hamaguchi2014a}, 
the best-fit model of the residual by a thermal ({\tt apec}) model ({\it dot}),
and the sum of these models ({\it solid grey}) are shown for comparison.
\label{fig:deepmin_spec}
}
\end{figure}

Since its launch in 2005, \SUZAKU\  has observed \etacar\ fourteen times.
Table~\ref{tbl:obslogs} summarizes the former ten observations, which were performed before 2011.
The first two observations were performed during the performance verification (PV) phase and their earlier result is summarized in \citet{Sekiguchi2009}.
The subsequent 8 observations were obtained through the guest observer program (AO-2, 3, 4, 6, PI: Kenji Hamaguchi).
Individual \SUZAKU\  observations are designated SUZ, subscripted with the year, month and day of the observation.

During these observations, \SUZAKU\ ran two types of instruments:
the X-ray Imaging Spectrometer \citep[XIS,][]{Koyama2006} in the focal plane of the thin-foil X-Ray Telescope \citep[XRT,][]{Serlemitsos2007}
and the Hard X-ray Detector \citep[HXD,][]{Takahashi2007,Kokubun2007}.
The XIS consists of four X-ray CCD cameras, XIS0$-$3, three of which (XIS0, 2 and 3)
use front-illuminated (FI) CCD chips, while one (XIS1) uses a back-illuminated (BI) chip.
The FI chips have good hard X-ray sensitivity, covering $\sim$0.5$-$10~keV,
while the BI chip has good soft X-ray sensitivity down to $\sim$0.3~keV.
The XIS2 was fatally damaged on 2006 Nov 9 by a mirco-meteorite, so this camera was unavailable after the 3rd observation (SUZ$_{070623}$).
Another micro-meteorite damaged one eighth of the XIS0 imaging area in 2009,
while multiple micro-meteorites probably produced small holes on optical blocking filters of all the XISs,
but these did not significantly degrade the data quality.
The XISs initially had good spectral resolution (FWHM $\sim$150~eV at 5.9~keV)\footnote{http://heasarc.gsfc.nasa.gov/docs/astroe/prop\_tools/suzaku\_td/node10.html}, 
but the resolution has gradually degraded with age due to radiation damage,
with a substantial recovery in 2006 October after initiating the Spaced Charge Injection (SCI) operation
with a sacrifice of the effective imaging area.
The XRT has a butterfly-shaped point spread function (PSF) with half power diameter (HPD) of $\sim$2$'$.
The effective area decreases as the off-axis angle increases, due to mirror vignetting.
The HXD consists of two types of detectors, the PIN with sensitivity between 15$-$70~keV and the GSO between 40$-$600~keV.
The GSO did not detect any significant signal above the non-X-ray background (NXB) level, so
we only used the PIN detector. The PIN detector has a collimator with a 34\ARCMIN$\times$34\ARCMIN\ \FOV,
on the bottom of which are PIN Si diodes.
The depletion voltage to the diodes has been reduced gradually to mitigate the increase of detector noise,
so that the detection efficiency has gradually decreased since launch.

The \SUZAKU\ point source observations have two default pointing positions --- the XIS nominal position,
which puts the main target at the XRT+XIS focus, and the HXD nominal position, which maximizes the
HXD collimator opening to the target.
The HXD nominal position is at 3.5\ARCMIN\ off-centered from the XIS nominal position.
In the PV observations, \etacar\ was placed at the XIS nominal position partly for instrument calibration.
In AO-2, 3 and 4, we placed \etacar\ at the HXD nominal position to maximize the HXD/PIN sensitivity to the star.
In AO-6, we put again \etacar\ at the XIS nominal position 
because a failure of a spacecraft gyro began to affect the XIS flux measurement at the HXD nominal position.
The satellite roll angles during the AO observations were optimized within the operational constraints
such that contamination from the nearby high energy sources AXP 1E~1048.1$-$5937 and IGR~J10447-6027 in the HXD/PIN \FOV\ are minimized ($\lesssim$1~\%, see the right panel of Figure~\ref{fig:obsimg}).
Only the HXD/PIN observation in SUZ$_{050829}$ included 5\% of the emission from 1E~1048.1$-$5937.

All the XIS observations were operated with the normal mode (no window option) 
because the count rates of \etacar\ for each XIS are $\lesssim$7~\UNITCPS, a factor of 2 below the threshold of significant photon pile-up.
However, the XIS pileup estimator \citep{Yamada2012} derived small pile-up for relatively high count rate observations
such as SUZ$_{081210}$ ($\sim$3\% pileup at the \PSF\ core).
Because of this artificial effect,
the XIS spectra in SUZ$_{080610}$ and SUZ$_{081210}$ significantly flatten above $\sim$9~keV.
We therefore excluded XIS spectra of these observations above 9~keV.
The XIS FI data had an anomaly at the first $\sim$9~ksec of SUZ$_{070623}$, whose interval we did not analyze.

\begin{figure*}[hdtp]
\epsscale{1.5}
\plotone{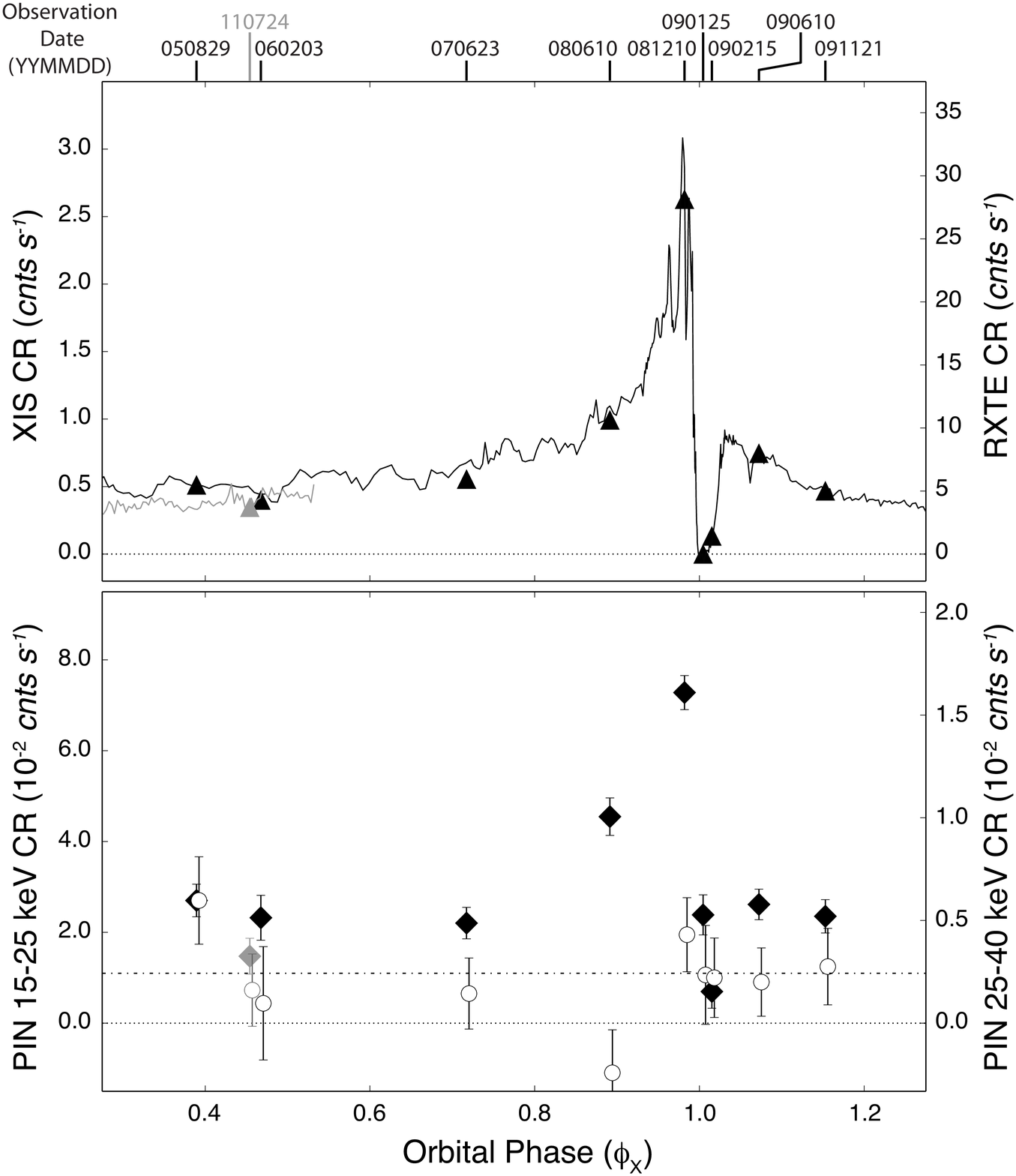}
\caption{{\it Top}: 
XIS light curve of the WWC X-rays between 5$-$9~keV (triangle), compared to the \RXTE\ light curves between 2$-$10 keV \citep[solid line,][]{Corcoran2010}.
The XIS0 and XIS3 count rates are normalized by the detector efficiency at SUZ$_{050829}$ and their count rates are averaged to derive the XIS count rate.
The fore- and back-ground emissions are estimated from the deep X-ray minimum data and subtracted, i.e.\ the count rate during the deep minimum is zero.
{\it Bottom}: 
HXD/PIN light curves between 15$-$25~keV (diamond) and 25$-$40 keV (open circle).
The 25-40 keV data points are slightly shifted to the right to show the error bars clearly.
The vertical bars show 1$\sigma$ errors, including the PIN systematic uncertainty of 1.3\%.
The dashed line shows the 25$-$40~keV flux of the \INTEGRAL\ point source in \citet{Leyder2008}.
In both panels, black and grey colors show intervals between 2005$-$2010 July and after 2010 August, respectively.
\label{fig:obstiming}
}
\end{figure*}

\begin{deluxetable}{l@{~~~~~~~~}cc@{~~~~~~~~}cc@{~~~~~~~~}cc@{~~~~~~~~}ccc@{~~~~~~~~}ccc}
\tablecolumns{13}
\tablewidth{0pc}
\tabletypesize{\scriptsize}
\tablecaption{Observed Count Rates\label{tbl:cnt_rate}}
\tablehead{
\multicolumn{1}{c}{Abbr}&
\multicolumn{2}{c}{XIS0~~~~~~}&
\multicolumn{2}{c}{XIS3~~~~~~}&
\multicolumn{2}{c}{XIS~~~~~~~~}&
\multicolumn{6}{c}{HXD}\\
\colhead{}&
\colhead{}&
\colhead{}&
\colhead{}&
\colhead{}&
\colhead{}&
\colhead{}&
\multicolumn{3}{c}{15$-$25~keV}&
\multicolumn{3}{c}{25$-$40~keV}\\
\colhead{}&
\multicolumn{1}{c}{CR}&
\multicolumn{1}{l}{Nor}&
\multicolumn{1}{c}{CR}&
\multicolumn{1}{l}{Nor}&
\multicolumn{1}{c}{CR}&
\multicolumn{1}{l}{Error}&
\multicolumn{1}{c}{CR}&
\multicolumn{1}{c}{Error}&
\multicolumn{1}{l}{Nor}&
\multicolumn{1}{c}{CR}&
\multicolumn{1}{c}{Error}&
\multicolumn{1}{c}{Nor}\\
\colhead{}&
\multicolumn{1}{c}{(cps)}&
\colhead{}&
\multicolumn{1}{c}{(cps)}&
\colhead{}&
\multicolumn{2}{c}{(cps)~~~~~~~~~}&
\multicolumn{2}{c}{(10$^{-2}$ cps)}&
\colhead{}&
\multicolumn{2}{c}{(10$^{-2}$ cps)}&
\colhead{}
}

\startdata
SUZ050829&0.56&1.00&0.56&1.00&0.558&0.002&2.70&0.36&1.00&0.60&0.21&1.00\\
SUZ060203&0.45&0.99&0.44&1.00&0.446&0.003&2.32&0.49&1.00&0.10&0.28&1.00\\
SUZ070623&0.63&0.72&0.57&0.86&0.601&0.003&2.20&0.35&1.05&0.14&0.17&1.04\\
SUZ080610&1.07&0.72&1.01&0.89&1.039&0.004&4.55&0.41&1.03&$-$0.24&0.21&1.04\\
SUZ081210&2.73&0.74&2.62&0.90&2.674&0.006&7.28&0.37&1.02&0.43&0.18&1.04\\
SUZ090125&0.05&0.72&0.04&0.89&0.047&0.001&2.39&0.44&1.02&0.24&0.24&1.04\\
SUZ090215&0.18&0.73&0.18&0.89&0.180&0.002&0.70&0.37&1.02&0.22&0.19&1.04\\
SUZ090610&0.81&0.72&0.78&0.89&0.792&0.003&2.62&0.34&1.02&0.20&0.17&1.04\\
SUZ091121&0.52&0.75&0.50&0.90&0.514&0.003&2.35&0.37&1.00&0.28&0.19&1.04\\
SUZ110724&0.39&0.92&0.39&0.95&0.392&0.002&1.47&0.40&0.88&0.16&0.18&0.95\\
\enddata

\tablecomments{
Abbr: Abbreviation adopted for each observation.
XIS0/XIS3: Net XIS0/XIS3 count rate (CR) between 5$-$9~keV, normalized with the effective area ratio to SUZ$_{050829}$ in the ``Nor" column.
XIS: Average of the XIS0 and XIS3 count rates.
HXD: Net HXD/PIN count rate between 15$-$25/25$-$40~keV, normalized with the effective area ratio to SUZ$_{050829}$ in the ``Nor" columns.
The contribution of the CXB and GRXE is estimated and subtracted.
The 1.3\% systematic error of the PIN NXB model is considered.
cps: \UNITCPS.
}

\end{deluxetable}

\subsection{Extraction of the WWC Emission Data}

In this paper, we analyze hard X-ray data above 5~keV to study the highest energy phenomena of \etacar.
For consistent analysis, we use data only from the XIS0 and XIS3 among the XISs,
both of which are FI sensors running through the \etacar\ observations.
We used the HEASoft version 6.14 and the CALDB version hxd20110913, xis20130724, and xrt20110630 for the data calibration.
The left panel of Figure~\ref{fig:obsimg} displays a 5$-$10~keV image from all the XIS3 data.
The brightest source at the center is \etacar, the second brightest source to the west of \etacar\ is the Wolf-Rayet (WR) binary system WR~25,
and the third to the north is the O4 star HD~93250.
The field includes more unresolved faint point sources, but no serendipitously bright X-ray source appeared during the observations in the XIS \FOV.

In the XIS analysis,
we defined a source region with a 2.5\ARCMIN\ radius circle centered at \etacar\ to minimize contamination from WR~25 and HD~93250.
The source region includes $\sim$90\% of X-ray photons from the star.
We extracted the background from an annulus with a 5\ARCMIN\ inner radius and a 8\ARCMIN\ outer radius centered at \etacar,
excluding areas within 3\ARCMIN\ from WR~25, HD~93250 and the centers of the X-ray clusters in \citet{Feigelson2011}.
The source region includes hard X-ray sources other than the central point source,
such as X-ray reflection at the HN \citep{Corcoran2004} and multiple young stars \citep[e.g.,][]{Wolk2011}.

\citet{Hamaguchi2014a} indicated that the WWC X-ray emission completely disappeared below 10~keV between 2009 Jan 12 and 28.
The SUZ$_{090125}$ observation was performed during this interval, so that the XIS spectrum should originate from the surrounding X-ray components.
The known hard X-ray components other than the WWC are the stable hot X-ray plasma, possibly in the foreground wind cavity \citep[the CCE component,][]{Hamaguchi2007b,Hamaguchi2014a},
which accounts for $\sim$55\% of the 5$-$10 keV emission, and X-ray reflection at the HN, which accounts for $\sim$33\% of the 5$-$10 keV emission (Figure~\ref{fig:deepmin_spec}).
The remaining $\sim$12\% probably originates from hard X-ray point sources within the source region.
The CCE component, which can be measured only during the X-ray minimum, did not vary more than $\sim$10\%
between 2003 and 2009 (Hamaguchi et al. in prep.), suggesting its stability over an orbital cycle.
The HN reflection emission is expected to decline by a factor of $\sim$4 from periastron (around SUZ$_{090125}$) to apastron (Hamaguchi et al. in prep.).
In this paper, we assume that the XIS data at SUZ$_{090125}$ represent the contaminating emission in all the XIS spectra.
However, this assumption significantly overestimates contribution of the fluorescent iron $K$ line from the HN to the apastron spectra
since the line flux around apastron is comparable to that at SUZ$_{090125}$.
We therefore defer the discussion of the fluorescent iron $K$ line emission to a later paper.

The HXD/PIN data include significant contamination from NXB, point sources,
Galactic Ridge X-ray Emission (GRXE) and cosmic ray background (CXB).
The NXB is estimated from the tuned background model with 1.3\% systematic uncertainty (1$\sigma$)
(JX-ISAS-SUZAKU-MEMO-2007-09\footnote{ftp://legacy.gsfc.nasa.gov/suzaku/data/background/pinnxb\_ver2.0\_tuned/}).
The only high-energy point source that could contaminate the HXD/PIN data is the AXP 1E~1048.1$-$5937.
However, the HXD/PIN count rate of 1E~1048.1$-$5937 on 2008 Nov.~30 (Obs ID: 403005010)
that excludes the NXB and the typical CXB spectrum was only 3.4$\times$10$^{-3}$~\UNITCPS (15$-$40~keV),
which is $\lesssim$1/5 of the \etacar\ count rate.
An extrapolation of the XIS spectrum to the HXD band accounts for only one-fifth of the detected HXD count rate:
the rest probably originates from GRXE.
In addition, the HXD band flux should not increase by more than a factor of two at any \SUZAKU\ observation of \etacar, 
considering a factor of $\lesssim$4 variation of this AXP in the soft band since 1996 \citep{Dib2014} and no strong color variation observed from the AXPs \citep{Enoto2010}.
Furthermore, $<$5\% of this AXP emission contributes to the HXD/PIN spectra of \etacar.
Considering all these results, contamination of this AXP, 1E~1048.1$-$5937, of the HXD spectra of \etacar\ should be negligible.
The GRXE emission around \etacar\ is estimated at 1.4$\times$10$^{-11}$~\UNITFLUX deg$^{-2}$ between 3$-$20~keV \cite[see section~5.2 in][]{Hamaguchi2007a}.
This is consistent with the remaining HXD/PIN spectrum of 1E~1048.1$-$5937, assuming an absorbed thermal plasma model ({\tt apec} $\times$ {\tt TBabs})
with \KT = 10 keV and \NH = 5.0$\times$10$^{22}$~\UNITNH.
Since the GRXE emission is not expected to vary strongly in 30\ARCMIN,
we use this spectrum as GRXE contamination in the HXD/PIN spectra of \etacar.
The CXB is estimated from the typical CXB emission \citep{Boldt1987}, which may 
fluctuate by $\lesssim$30\% from region to region \citep{Miyaji1998}.
The CXB/GRXE contributions are estimated at 1.4$\times$10$^{-2}$/4.1$\times$10$^{-3}$~\UNITCPS\ [15$-$25~keV] and 
3.7$\times$10$^{-3}$/4.8$\times$10$^{-4}$~\UNITCPS\ [25$-$40~keV], while 
the CCE and HN contribution should be negligible ($\lesssim$10$^{-3}$~\UNITCPS\ [15$-$25~keV], $\lesssim$10$^{-4}$~\UNITCPS\ [25$-$40~keV]).
The CXB and GRXE contribution is excluded from the PIN count rates and spectra as background.
In the light curve analysis, 
we assume that the statistical noise errors are Gaussian and assume a systematic uncertainty of 1.3\% in the HXD/PIN NXB model.

The detector effective areas to \etacar\ varied by up to $\sim$30\% for the XIS and $\sim$12\% for the HXD/PIN between the observations
because of changes in observing conditions ---
the nominal pointing position, the XIS SCI operation and the sensitivity degradation of the PIN sensors.
The efficiency variation in the spectral analysis is automatically considered with spectral responses generated with {\tt xisrmfgen} and {\tt xisarfgen} in the HEAsoft tools for 
the XIS and provided by the calibration team through the calibration database\footnote{http://heasarc.gsfc.nasa.gov/docs/heasarc/caldb/suzaku/} for the HXD.
In the light curve analysis, the average efficiency in a given energy band is calculated from the generated spectral response,
and the count rates of each observation are normalized at the detector efficiency at SUZ$_{050829}$.

\begin{figure*}[t]
\epsscale{1.3}
\plotone{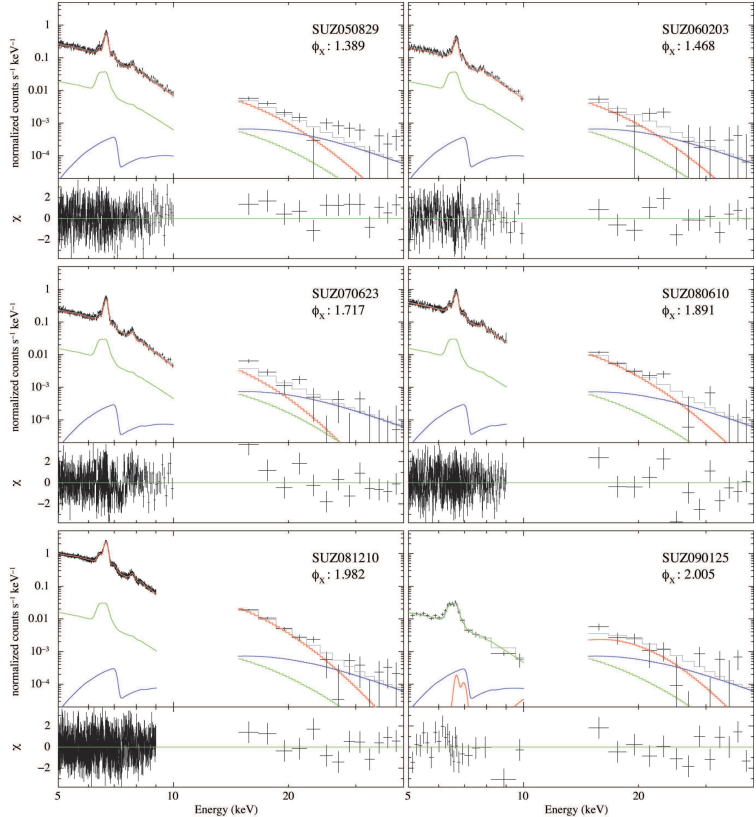}
\caption{XIS and HXD spectra of \etacar\ and the best-fit model in Tables~\ref{tbl:specfit_variable}, \ref{tbl:specfit_constant}.
The HXD spectra exclude expected contributions from CXB and GRXE, i.e.\ both the XIS and HXD spectra should originate within 2.5\ARCMIN\ of \etacar.
The red, blue, green and grey lines in each panel represent the WWC thermal component, the power-law component,
the stable foreground thermal component (i.e.\ CCE, HN and surrounding point sources), and their total, respectively.
Each bottom panel shows the residuals of the $\chi^{2}$ fit.
\label{fig:obsspec}
}
\end{figure*}

\begin{figure*}[t]
\figurenum{4}
\plotone{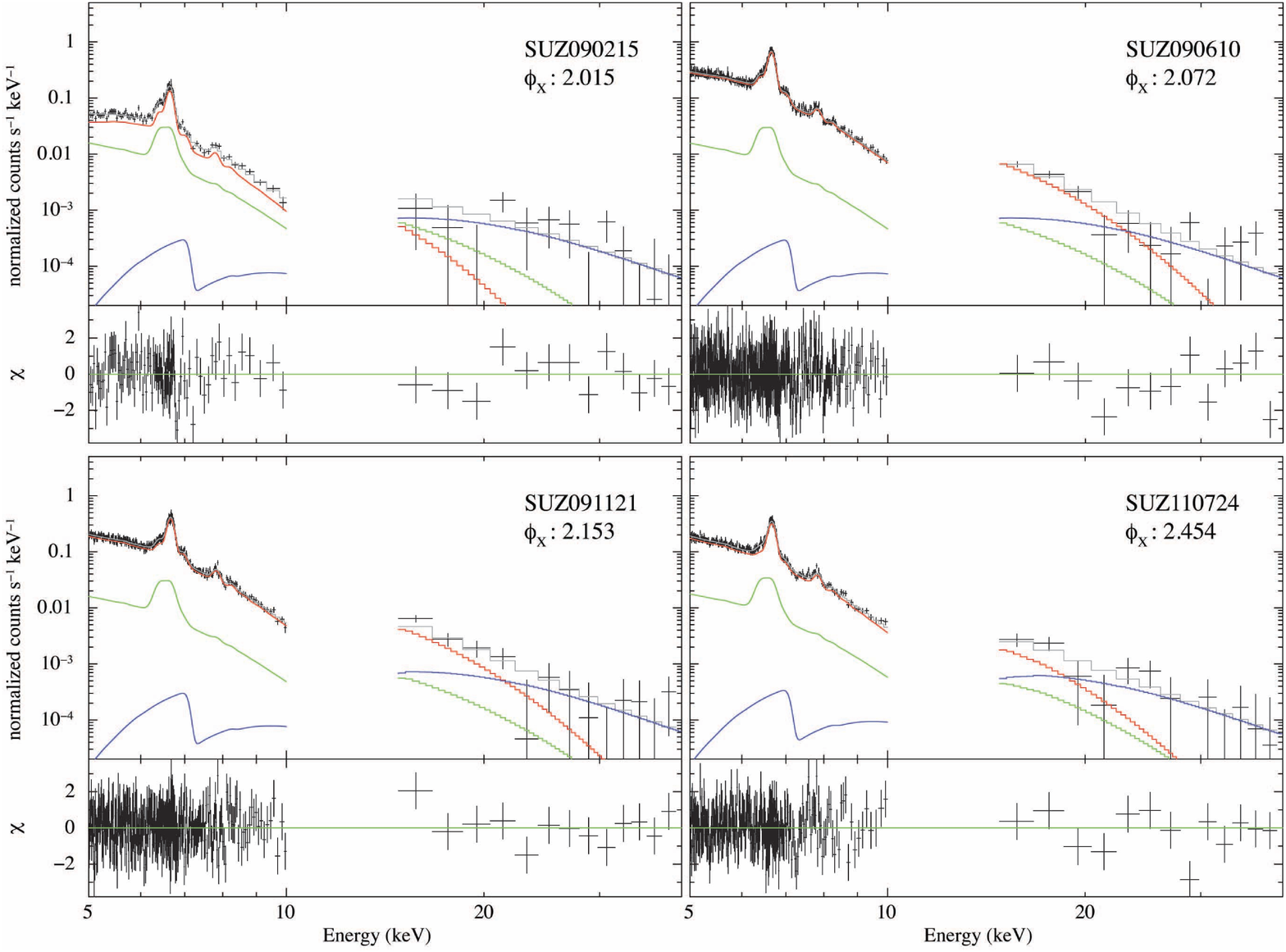}
\caption{Continued.
\label{fig:obsspec1}
}
\end{figure*}

\section{Result}

\subsection{Light curve}

For each observation,
we produced an XIS light curve between 5$-$9~keV and an HXD/PIN light curve between 15$-$25~keV with 500~sec time bins.
The XIS light curves at SUZ$_{050829}$, SUZ$_{081210}$, SUZ$_{090610}$, SUZ$_{091121}$, and SUZ$_{110724}$
with good photon statistics reject a constant model at a confidence limit above 95\%.
These variations are apparently caused by small flux fluctuations on timescales of $\lesssim$2~ksec and not by a systematic variation on long timescales,
as seen in the \XMM\ observations in 2003 \citep{Hamaguchi2007b}.
The HXD/PIN light curves at SUZ$_{050829}$, SUZ$_{060203}$, and SUZ$_{070623}$ reject a constant model at above 95\% confidence, 
but they did not show any apparent long-term variation, either.

We measured from each observation the net XIS count rate between 5$-$9~keV 
and the net HXD/PIN count rates between 15$-$25~keV and 25$-$40~keV (Table~\ref{tbl:cnt_rate}).
The top panel of Figure~\ref{fig:obstiming} shows the XIS light curve between 5$-$9~keV.
In this plot, we subtracted the XIS count rate in SUZ$_{090125}$ (0.05~\UNITCPS) from 
XIS count rates of all the XIS observations as contamination from surrounding X-ray sources.
We also compare this to the \RXTE\ light curve after 2005 between 2$-$10~keV \citep{Corcoran2010}.
The amount of contamination of X-ray sources in the \RXTE\ \FOV\ is also estimated from the 
deep X-ray minimum observations \citep[see details in][]{Hamaguchi2014a} and subtracted from the \RXTE\ light curve.
This means that both light curves should be of the WWC X-ray emission and directly comparable.
Their vertical axes are scaled such that the same height gives the same energy flux
in the typical \etacar\ spectrum (\KT\ $\sim$4.5~keV, \NH\ $\sim$5$\times$10$^{22}$~\UNITNH\ and $Z \sim$0.8~solar).
These two light curves match very well.

The bottom panel of Figure~\ref{fig:obstiming} shows the HXD/PIN light curves between 15$-$25~keV and 25$-$40~keV.
Their vertical axes are scaled such that the data points at SUZ$_{050829}$ overlap.
The 15$-$25~keV light curve varied similarly to the 5$-$9~keV light curve outside the minimum: it increased gradually toward periastron.
This result indicates that the 15$-$25~keV emission has the same origin as the 5$-$10~keV emission, i.e.\ the WWC X-rays.
It, however, varied differently during the minimum.
It declined only by a factor of 3 from SUZ$_{081210}$ at SUZ$_{090125}$ when the 5$-$9~keV flux dropped to zero.
The minimum observed flux occurred during the next observation (SUZ$_{090215}$) during the shallow minimum phase.

On the other hand, the 25$-$40~keV light curve did not show any significant variation near the X-ray maximum,
accepting a constant flux model (reduced $\chi^{2}=$ 1.11 for d.o.f. = 9).
We converted the 22$-$100~keV flux of the \INTEGRAL\ source \citep[][0.15~\UNITCPS]{Leyder2008} to the HXD/PIN count rate 
between 25$-$40~keV, assuming a $\Gamma =1.1$ power-law spectrum, and plotted it with a dotted line.
This flux level matches quite well all the data points except SUZ$_{050829}$ and SUZ$_{080610}$.
The result suggests that the 25$-$40~keV emission does not originate from the WWC thermal plasma, but from 
the same component as the \INTEGRAL\ source.

In summary, the light curve analysis suggests two major X-ray emission components between 5$-$40~keV:
strongly variable emission below $\sim$25~keV and stable emission above $\sim$25~keV.
The former component probably corresponds to the WWC thermal plasma emission and the latter to the power-law component 
\citep{Leyder2008,Sekiguchi2009,Leyder2010}.

\begin{figure*}[t]
\epsscale{1.5}
\plotone{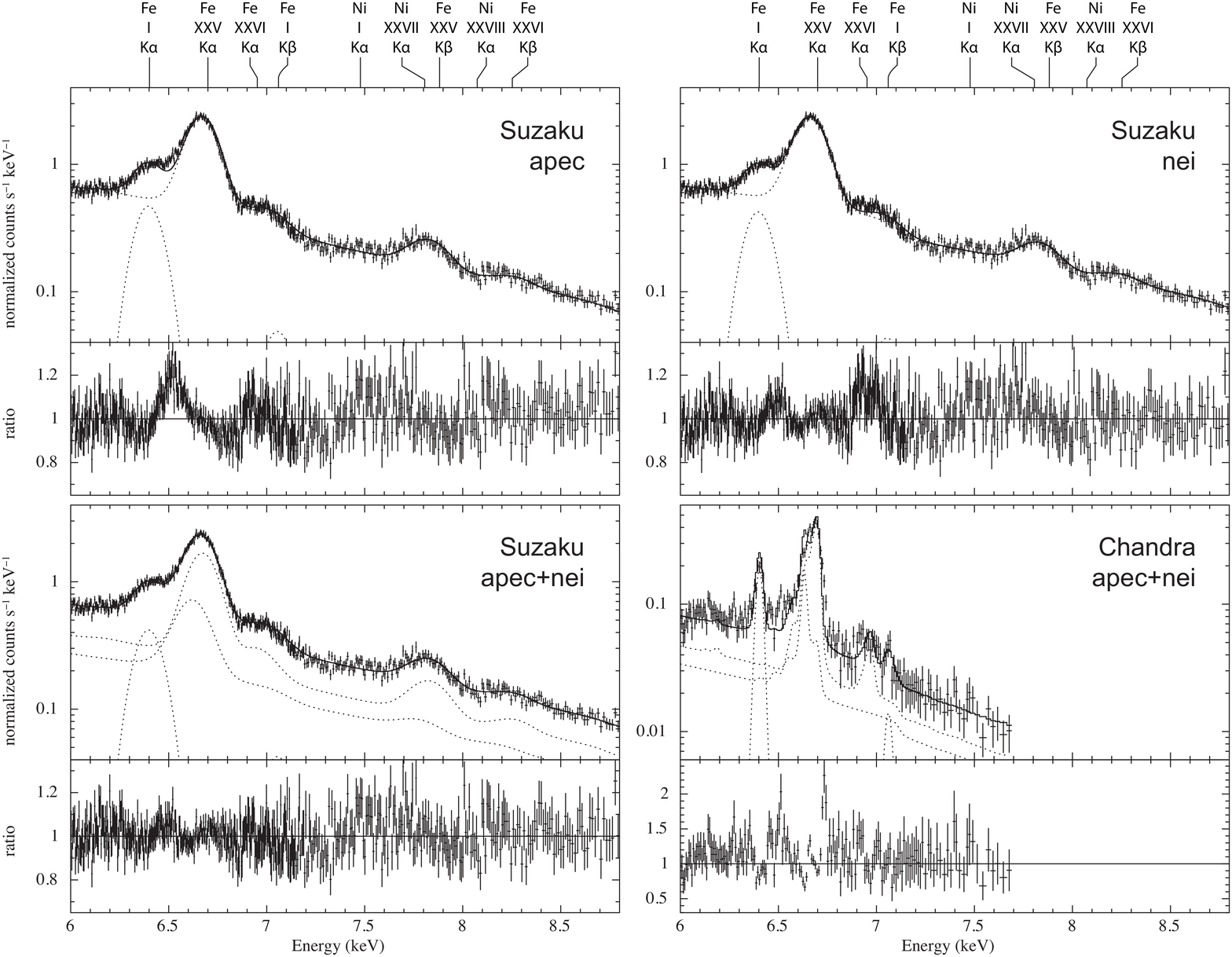}
\caption{
XIS spectrum in SUZ$_{081210}$ fit by an absorbed CE plasma model ({\tt apec}, {\it top left}), 
an absorbed NEI plasma model ({\tt nei}, {\it top right})
and an absorbed CE + NEI plasma model ({\tt apec}+{\tt nei}, {\it bottom left}),
and the \CHANDRA\ HETG spectrum obtained between 2008 December 8$-$13 overlaying the best-fit {\tt apec}+{\tt nei} model for the SUZ$_{081210}$ spectrum ({\it bottom right}).
The \CHANDRA\ spectrum above $\sim$7.75~keV is contaminated by the METG grating data and therefore is not extracted.
The model for the \CHANDRA\ spectrum includes the CCE component, but not the HN nor the surrounding point source components,
which are outside of the \CHANDRA\ event extracting region.
\label{fig:xis_spec_issue}
}
\end{figure*}

\subsection{Spectrum}

Figures~\ref{fig:obsspec} shows the XIS0+3 and HXD/PIN spectra of all observations.
The XIS spectra varied as in the 2003 orbital cycle \citep{Hamaguchi2007b};
the hard band slope between 7$-$10~keV, which reflects the hottest temperature of the WWC plasma,
did not vary significantly through the cycle.
The HXD spectra below 25~keV seem to connect smoothly to the XIS hard band.
The spectra above 25~keV do not show prominent features within the limited photon statistics.

The Helium-like iron line complex at $\sim$6.7~keV distorted strongly toward the low energy side
around periastron, as seen in the previous cycle \citep{Hamaguchi2007b}.
To show this distortion clearly,
we first fit the XIS spectrum in SUZ$_{081210}$ above 5~keV by an absorbed 1$T$ collision equilibrium (CE) plasma model 
({\tt apec}, {\it top left} panel of Figure~\ref{fig:xis_spec_issue}).
The spectrum also shows a fluorescent line from cold iron at 6.4 keV, for which we assume a narrow Gaussian line,
based on a \CHANDRA\ grating observation of \etacar\ around apastron in 2000 \citep{Corcoran2001a}.
We also add another narrow Gaussian line for Fe K$\beta$ fluorescence at 7.06~keV with the intensity tied to 12.2\% of the Fe K$\alpha$ line \citep{Yamaguchi2014a}.
The spectrum also shows an iron $K$ absorption edge at 7.1~keV, for which the column density of cold iron (\NFE) is varied independently
from the hydrogen column density (\NH).
The best-fit model has a strong excess at $\sim$6.5~keV, for which the 6.4~keV line is overestimated to compensate --- 
this result suggests more emission from lowly ionized iron.
A marginal enhancement between Ni I $K\alpha$ and Ni XXVII $K\alpha$ lines also support the presence of lowly ionized nickel.

However, the {\tt nei} model, a non-equilibrium ionization (NEI) plasma model in {\tt xspec} (NEIVERS version 3.0),
does not reproduce the Hydrogen-like iron line at 6.9 keV in the spectrum ({\it top right} panel of Figure~\ref{fig:xis_spec_issue}).
This model still does not reproduce the excess at $\sim$6.5~keV.
The {\tt pshock} model, which considers plasma distribution in different ionization timescales in the plane parallel shock, gives a similar result.
Thus, there has to be a significant amount of CE plasma emission to reproduce the Hydrogen-like lines.

We, therefore, fit this spectrum by an absorbed {\tt apec} plus {\tt nei} model as a testbed.
The plasma temperatures of the {\tt apec} and {\tt nei} components cannot be independently determined and therefore are tied.
Their elemental abundances are also tied together.
The best-fit model reproduces the XIS spectrum well ({\it bottom left} panel of Figure~\ref{fig:xis_spec_issue}).
The excess at $\sim$6.5~keV totally disappears, while the Hydrogen-lines are reproduced well.
This best-fit model also reproduces well the \CHANDRA\ HETG grating spectrum obtained quasi simultaneously
to SUZ$_{081210}$ between 2009 Dec 8$-$13
({\it bottom right} panel of Figure~\ref{fig:xis_spec_issue}, Observation ID: 10831, 8930, 10827, Total exposure: 74.6~ksec, PI: Corcoran, M.~F.).
With a factor of $\sim$5 better spectral resolution than the XISs, the NEI component is clearly seen as a red wing of the Helium-like Fe $K\alpha$ line.
The residual at the blue side of the Helium-like Fe $K\alpha$ line, which cannot be resolved with the CCD resolution,
can be reproduced by a Doppler broadening of $\Delta v\sim$800~\UNITVEL.

We therefore use this model for all the spectra but SUZ$_{090125}$, which does not show WWC emission below 10~keV.
The spectra outside of the X-ray minimum and maximum (SUZ$_{050829}$, SUZ$_{060203}$, SUZ$_{070623}$, SUZ$_{091121}$ and SUZ$_{110724}$) 
do not show clear distortions in the Fe $K\alpha$ line but small excesses at $\sim$6.5~keV, so that we tie their ionization parameters.
The elemental abundances of all the emission components are tied together and the abundance ratios between elements
above Helium are fixed at the values of the {\tt aspl} solar abundance model \citep{Asplund2009} in the XSPEC modeling.
This is a reasonable approach because the hot plasma is heated by the secondary stellar winds and so is expected to 
reflect the elemental abundance of the secondary star, which has an unknown evolutionary status.
We also assume the elemental abundance of the absorber to be solar except for iron.
The absorbing material should originate from the primary star, which is depleted in hydrogen, carbon and oxygen but rich in nitrogen.
However, these elements do not affect the spectral structure above 5~keV.
The other high-$Z$ elements are considered to be solar \citep{Hillier2001}.
The absolute \NH\ depends on the hydrogen abundance, but it can be easily adjusted for any abundance models later.

To this variable WWC model, we add an absorbed power-law model for the extremely hard X-ray component above $\sim$25~keV.
The HXD/PIN spectra between 25$-$40 keV do not have enough statistics to determine the power-law index and normalization, individually.
We therefore fixed the power-law index at 1.4.
The result does not change substantially for power law indices in the range 1.0$-$1.8.
Since the 25$-$40~keV light curve did not show significant flux variations,
we tied the normalization of the power-law component between observations.
\citet{Leyder2010} constrained the position of this power-law source to within 1.6\ARCMIN\ from \etacar\ from the \INTEGRAL\ observations.
Since the XIS source region is within 2.5\ARCMIN\ from \etacar, the XIS spectra should also include this power-law source.
A simple extrapolation of this $\Gamma\sim$1.0$-$1.8 power-law spectrum should show significant emission below $\sim$10~keV during the deep X-ray minimum,
but neither the XIS spectrum at SUZ$_{090125}$, nor any \CHANDRA, nor \XMM\ spectra during the minimum suggest the presence of this power-law source \citep{Hamaguchi2014a}.
\citet{Leyder2010} did not find any promising candidate of this counterpart other than \etacar\ in a \CHANDRA\ image, either.
This means that this power-law component is heavily absorbed, at least during the X-ray minimum, and cannot be seen below 10~keV.
The absorption to this power-law component cannot be constrained outside the X-ray minimum.
We, therefore, assume a constant absorption to this power-law component through the orbital cycle.

\begin{deluxetable}{llllllll}
\rotate
\tabletypesize{\scriptsize}
\tablecolumns{8}
\tablewidth{0pc}
\tablecaption{Best Fit Spectral Model --- Variable WWC Parameters\label{tbl:specfit_variable}}
\tablehead{
\colhead{Abbr}&
\colhead{\KT}&
\colhead{{\it E.M.}\ [{\tt apec}]}&
\colhead{{\it E.M.}\ [{\tt nei}]}&
\colhead{$\tau$~[{\tt nei}]}&
\colhead{Flux (6.4 keV)}&
\colhead{\NH}&
\colhead{$N_{\rm Fe}$}\\
\colhead{}&
\colhead{(keV)}&
\colhead{(10$^{57}$~cm$^{-3}$)}&
\colhead{(10$^{57}$~cm$^{-3}$)}&
\colhead{(10$^{10}$~cm$^{-3}$~s$^{-1}$)}&
\colhead{(10$^{-4}$~ph~cm$^{-2}$~s$^{-1}$)}&
\colhead{(10$^{23}$~\UNITNH)}&
\colhead{(10$^{23}$~\UNITNH)}
}
\startdata
SUZ050829&4.5~(4.4,4.7)&3.9~(3.6,4.2)&1.3~(1.0,1.4)&6.2~(4.8,8.7)&0.41~(0.35,0.47)$^{\dagger}$&0.21~($<$0.44)&1.2~(0.84,1.6)\\
SUZ060203&4.9~(4.6,5.2)&2.6~(2.4,2.9)&1.1~(0.8,1.4)&=$\tau_{\rm 050829}$&0.20~(0.11,0.28)$^{\dagger}$&0.0~($<$2.5)&1.5~(0.89,2.2)\\
SUZ070623&3.3~(3.0,3.5)&6.0~(5.0,7.2)&3.6~(2.9,4.5)&=$\tau_{\rm 050829}$&0.41~(0.34,0.48)$^{\dagger}$&1.2~(0.79,1.6)&1.6~(1.1,2.0)\\
SUZ080610&4.6~(4.3,5.0)&6.0~(4.7,7.0)&3.9~(3.4,4.5)&8.2~(6.4,11.1)&1.2~(1.1,1.3)&0.43~(0.07,0.77)&1.5~(1.1,1.9)\\
SUZ081210&3.7~(3.6,3.8)&23.5~(22.5,25.4)&16.5~(15.3,17.8)&7.8~(6.6,9.6)&5.1~(4.9,5.3)&1.6~(1.4,1.8)&2.2~(2.0,2.4)\\
SUZ090125&4.0~(fix)&33.4~(16.7,61.9)&\nodata&\nodata&\nodata&82.5~(63.8,125.8)&=\NH$_{\rm 090125}$\\
SUZ090215&2.3~(1.9,2.9)&3.0~($<$8.3)&6.0~(3.9,9.7)&12.8~(9.2,17.3)&0.54~(0.42,0.60)&6.1~(4.6,6.8)&4.7~(3.3,6.0)\\
SUZ090610&4.1~(3.8,4.4)&2.6~(1.6,3.7)&7.4~(6.2,8.7)&10.5~(9.5,11.6)&1.0~(0.90,1.1)&1.2~(0.82,1.5)&3.2~(2.8,3.6)\\
SUZ091121&4.3~(4.1,4.5)&3.6~(3.4,4.0)&1.7~(1.3,2.1)&=$\tau_{\rm 050829}$&0.45~(0.37,0.52)&0.67~(0.39,1.0)&1.6~(1.1,2.1)\\
SUZ110724&3.8~(3.6,4.0)&3.0~(2.8,3.3)&1.1~(0.89,1.3)&=$\tau_{\rm 050829}$&0.17~(0.10,0.23)$^{\dagger}$&0.0~($<$3.1)&0.87~(0.33,1.4)\\
\enddata
\tablecomments{
Spectral Model: WWC + power-law + (CCE + HN + surrounding point sources).
The WWC component is a combination of the models ({\tt apec} + {\tt nei} + {\tt Gaussian}[6.4~keV] + {\tt Gaussian}[7.1~keV]) $\times$ {\tt varabs},
while the power-law component is {\tt powerlaw} $\times$ {\tt TBabs}.
The best-fit result of the common parameters --- elemental abundance of the WWC component and normalization and absorption of the power-law component ---
is separately shown in Table~\ref{tbl:specfit_constant}.
The parentheses show the 90\% confidence ranges.
If not specified, the model assumes the elemental abundance relative to hydrogen in the {\tt aspl} solar abundance model.
$^{\dagger}$These numbers should significantly underestimate the line fluxes from the WWC vicinity because this model does not consider variation of the HN reflection emission
through the orbital cycle.
}
\end{deluxetable}

\begin{deluxetable}{ccc}
\tabletypesize{\scriptsize}
\tablecolumns{3}
\tablewidth{0pc}
\tablecaption{Best Fit Spectral Model --- Constant Parameters \label{tbl:specfit_constant}}
\tablehead{
\colhead{Abundance}&
\multicolumn{2}{c}{Power-Law}\\
\colhead{}&
\colhead{Normalization}&
\colhead{\NH}\\
\colhead{(solar)}&
\colhead{(10$^{-4}$)}&
\colhead{(10$^{24}$~\UNITNH)}
}
\startdata
0.91~(0.85,0.98)&3.5~(2.7,4.5)&2.4~(1.5,3.9)\\
\enddata
\tablecomments{
See Table~\ref{tbl:specfit_variable} for details.
}
\end{deluxetable}

The 15$-$25~keV flux at SUZ$_{090125}$ is too high for either the thermal component seen in the XIS band 
or the power-law component above 25~keV.
We therefore assume the excess as the deeply embedded WWC emission and reproduce it with the model for the WWC thermal component.
Since the statistics are limited, we fixed the plasma temperature at 4~keV, the typical temperature of the WWC plasma outside of the X-ray minima.
We also tied \NH\ and \NFE\ of SUZ$_{090125}$ because the Fe $K$ absorption edge cannot be measured.
To all the spectral models, we add the best-fit model of the XIS spectrum in SUZ$_{090125}$ 
to account for emission from the CCE, the HN and the surrounding point sources.

In this model fit, the model normalizations for the HXD/PIN spectra were multiplied by 1.15 for the XIS nominal pointing observations
and 1.19 for the HXD nominal pointing observations,
according to the Suzaku Data Reduction Guide\footnote{http://heasarc.gsfc.nasa.gov/docs/suzaku/analysis/abc/}.

The best-fit model reproduced all the spectra very well (reduced $\chi^2$ =1.08, d.o.f =3437, Tables~\ref{tbl:specfit_variable}, \ref{tbl:specfit_constant} and Figure~\ref{fig:obsspec}).
The hottest plasma temperatures of the thermal WWC component are stable at \KT\ $\sim$4~keV outside the minimum
and the HXD/PIN spectra showed no signature of hotter plasma.
The fit to the SUZ$_{090215}$ spectrum resulted in a low plasma temperature of $\sim$2~keV, similarly to spectral fits
to the shallow minimum spectra in 2003 \citep{Hamaguchi2007b}.
However, \KT\ is degenerate with \NH\ in fits to strongly absorbed spectra,
so that this variation may not suggest an actual decline in temperature of the hottest plasma.
The elemental abundance of the plasma is close to solar.
The \NH\ goes down to zero for SUZ$_{060203}$ and SUZ$_{110724}$.
However, the extrapolations of these models significantly overestimate the spectra below 5~keV;
the soft band spectra suggest higher \NH~$\sim$3$-$5$\times$10$^{22}$~\UNITNH.
The \NH\ probably does not correctly represent the absorption column to the hot plasma 
because of lower temperature plasma emission important around 5~keV.
On the other hand, \NFE\ is determined from the iron edge feature and is therefore a more reliable estimator of the absorption to the hot plasma.
The \NFE\ increased toward the deep minimum and reached the maximum of $\approx$8$\times$10$^{24}$~\UNITNH\ at SUZ$_{090125}$.
The power-law component also required a very high value of \NH~$\sim$2$\times$10$^{24}$~\UNITNH.

\section{Discussion}

\subsection{Orbital Modulation of the Physical Parameters}

\SUZAKU\ sampled a whole orbital cycle of \etacar\ between 2005 and 2011.
Though the X-ray minimum in 2009 was significantly shorter than those in previous cycles \citep{Corcoran2010},
the latest \SUZAKU\ spectrum was very similar to that in the previous cycle, again suggesting a cyclic variation.

\begin{figure*}[t]
\plotone{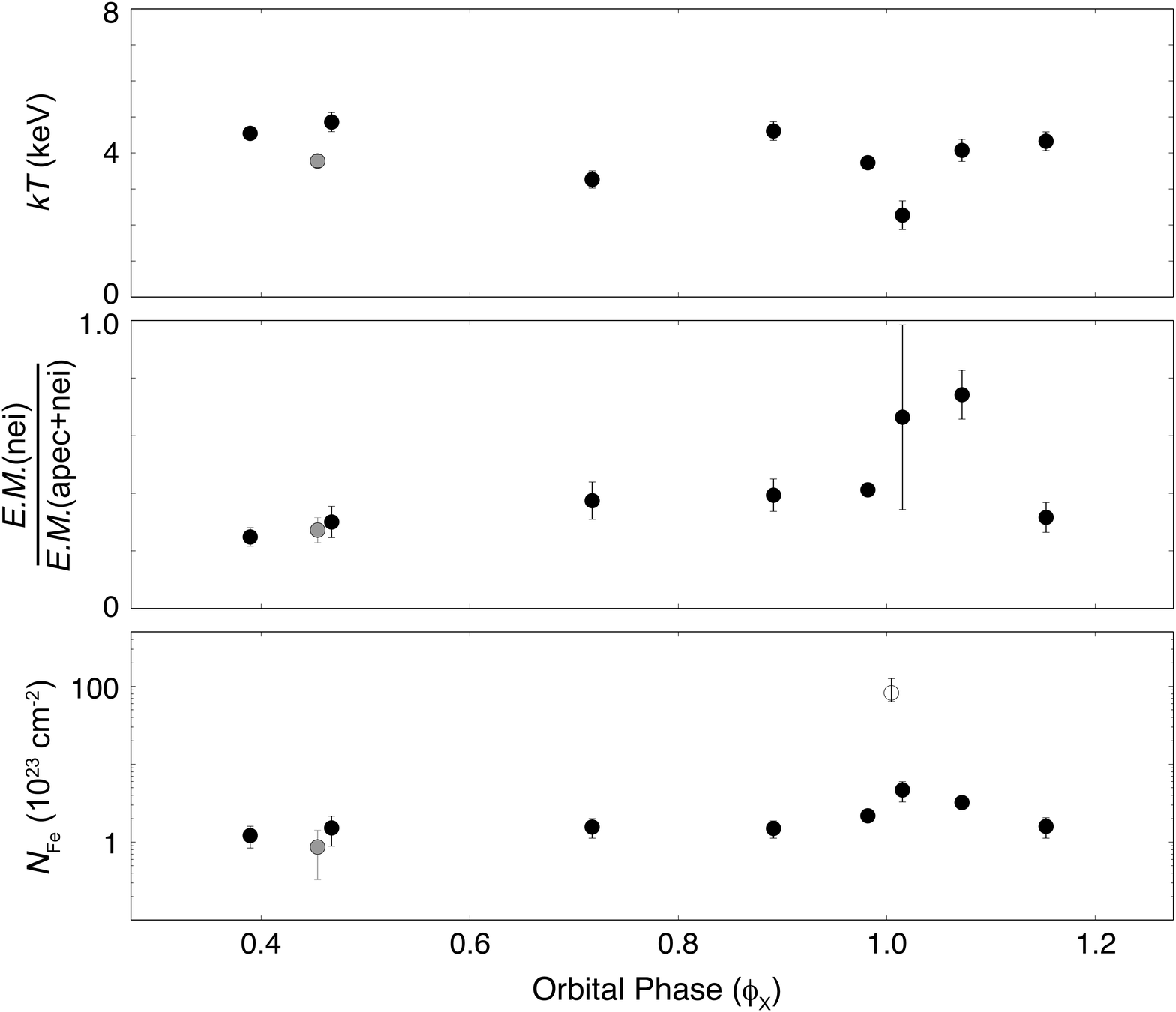}
\caption{Variations of the selected spectral parameters in the best-fit model 
({\it Top}: plasma temperature, {\it Middle}: emission measure ratio of the NEI ({\tt nei}) plasma,
{\it Bottom}: absorption column density of cold iron in units of an equivalent hydrogen column density
in the case of solar abundance matter.
The open circle at the bottom panel is of SUZ$_{090125}$, measured from the soft band cut-off.
The vertical bars depict 90\% error ranges.
\label{fig:spec_param}
}
\end{figure*}

The combined fit to the XIS and HXD/PIN spectra confirmed that the hottest temperatures of the WWC plasma
are stable at \KT~$\sim$4$-$5~keV through the orbit outside the X-ray minimum (Figure~\ref{fig:spec_param} {\it top}).
The same conclusion was deduced by \citet{Ishibashi1999} using the \RXTE\ data obtained between 1996 April and 1998 October
and \citet{Hamaguchi2007b} using the \XMM\ and \CHANDRA\ data obtained between 2000 July and 2003 September.
The new \SUZAKU\ result is important in two ways.
The set of observations has a long baseline between 2005 August and 2011 July and samples the whole orbital cycle.
The HXD/PIN provided the best quality measure of the extremely high energy spectra above 15~keV with a smaller \FOV\ and
lower background than the \RXTE\ Proportional Counter Array (PCA).
The wide band coverage realized with the HXD provided a better measurement
of the continuum slope in the very high energy range, and therefore increased the sensitivity to the hotter temperature plasma.
The \SUZAKU\ results show that the plasma temperature does not change prominently outside the minimum,
in line with predictions by WWC theories.

The distortion of the Helium-like Fe $K\alpha$ line was first recognized during the X-ray minimum in 2003
and discussed as caused by the NEI effect \citep{Hamaguchi2007b}.
The new \SUZAKU\ result demonstrates that an NEI plasma with $\tau~\sim$0.6$-$1.3$\times$10$^{11}$~cm$^{-3}$~s$^{-1}$
can reproduce this distortion in spectral fits and 
this NEI plasma component can be present through the orbital cycle, with an increased ratio around periastron.
The {\tt pshock} model, which considers the ionization timescale distribution in a plane parallel shock,
fails to reproduce the whole He-like Fe $K\alpha$ line profile.
This result suggests that the plasma really has two peaks in the ionization timescale distribution at
$\sim$10$^{11}$~cm$^{-3}$~s$^{-1}$ (the NEI component) and above $\sim$10$^{12}$~cm$^{-3}$~s$^{-1}$ (the CE component).
This result probably means that the NEI plasma heats up in $\sim$1000 (10$^{8}$~\UNITPPCC$/n$) sec, where $n$ is the plasma density,
and then quickly cools down without reaching the thermal equilibrium.
The \EM\ ratio of the NEI plasma increases from $\sim$25\% around apastron to $\sim$75\% around periastron (Figure~\ref{fig:spec_param} {\it middle}).
The ratio is high with a large uncertainty during the shallow minimum (SUZ$_{090215}$), 
but it is also high after the recovery (SUZ$_{090610}$) as well.
This result does not suggest that the ratio is correlated with the X-ray luminosity.
The SUZ$_{090610}$ observation is when the WWC apex is wrapped inside the primary wind, and therefore in a high density environment.
The NEI plasma may quickly contact the thick cool primary wind and cool down, while the CE plasma may rapidly leave the system in the
part of the shocked secondary wind which does not come into direct contact with the higher density primary wind.

The absorption column density of cold iron (\NFE) increases by a factor of 2$-$4 toward periastron (Figure~\ref{fig:spec_param} {\it bottom}).
This variation is very similar to the \NH\ variation measured from hard band spectra above 5 keV in 2003 \citep{Hamaguchi2007b}; 
the absorption to the WWC apex varies periodically as well.
Interestingly, the lowest \NFE\ observed around apastron is still a factor of 2$-$3 higher than absorption to soft X-rays ($\sim$4$-$5$\times$10$^{22}$~\UNITNH),
though the WWC apex should be seen through the thin secondary wind.
Since there is no evidence of iron overabundance in \etacar\ \citep[e.g.,][]{Hamaguchi2007b,Hillier2001},
this perhaps could be explained if the secondary wind piles up over the WWC contact surface.

\subsection{High 15$-$25~keV Flux during the Deep X-ray Minimum}

The relatively strong 15$-$25~keV emission at SUZ$_{090125}$ can be reproduced by the WWC emission
viewed through extremely high photoelectric absorption (\NH\ $\approx$8$\times$10$^{24}$~\UNITNH).
The \CHANDRA\ spectrum obtained at the end of the X-ray eclipse on 2009 Feb 3 also suggested 
a very high \NH\ of $\sim$10$^{24}$~\UNITNH\ \citep{Hamaguchi2014a}, 
and the WWC apex should be more embedded at the middle of the X-ray eclipse.
A peak \NH\ of $\sim$several$\times$10$^{24}$~\UNITNH\ during periastron is also suggested by simulations of WWC X-ray emission from \etacar\ 
\citep[][Russell et al.\ in preparation]{Parkin2011}.
The large \NH\ at SUZ$_{090125}$ is consistent with the picture that 
the WWC X-ray emission peered through very thick intervening material that totally blocked X-ray emission below 10 keV.
If this interpretation is correct,
the intervening material would be the inner primary wind, and not the primary stellar body.

In this interpretation, the WWC activity during the deep X-ray minimum is still strong behind the absorber.
The \EM\ at SUZ$_{090125}$ is as large as that at the maximum in SUZ$_{081210}$.
This is consistent with the WWC theory, in which the luminosity is inversely proportional to the distance between the two stars.
However, the Compton scattering process becomes important at this absorption column.
Since emission scattered off the line of sight may end up reaching us after another scattering,
the amount of attenuation by the Compton scattering also depends on the shape of the surrounding 
intervening material.
A broad-band spectrum above 10~keV at this phase with good photon statistics is required to correctly 
measure the amount of the Compton scattering and hence the intrinsic luminosity.

\subsection{Origin of the Power-law Component}
The PIN count rates between 25$-$40~keV did not vary strongly.
The $\gamma$-ray source, 1FGL~J1045.2-5942, detected by the \FERMI\ $\gamma$-ray observatory also only varied by a factor of $\lesssim$2
including the X-ray minimum, with a possible weak decline after the recovery \citep{Abdo2010,Farnier2011,Reitberger2012}.
This result strengthens the hypothesis 
that the 25$-$40~keV power-law source is connected to the \FERMI\ $\gamma$-ray source \citep{Leyder2008,Leyder2010,Abdo2010}.

In this interpretation, the power-law component originates from emission up-scattered by GeV particles accelerated at the WWC region.
However, our results show that this power-law component does not change significantly around the maximum when the WWC head-on
collision is the strongest and around shallow minimum when emission near the WWC apex apparently shuts off \citep[][and additional references therein]{Hamaguchi2014a}.
Our result does not suggest that the power-law component does not originate near the WWC stagnation point.
It remains to be seen whether the power-law component can be reproduced by models of particle acceleration in the WWC.

To satisfy low X-ray flux below 10~keV of \etacar\ during the deep X-ray minimum,
the power-law source should suffer extremely strong absorption of \NH~$\approx$2$\times$10$^{24}$~\UNITNH.
This high \NH\ does not favor the foreground shock region such as the CCE plasma cavity nor the HN lobe,
whose extinctions are less than \NH~$\sim$5$\times$10$^{22}$~\UNITNH.
One obvious but less interesting hypothesis is an unrelated neutron star or an active galactic nuclei behind \etacar,
though the chance of this coincidence is not high.
A provocative but more interesting hypothesis may be the presence of an active compact object associated with the binary system,
which was once bound to the binary system but ejected by events such as the 1840 eruption. 
The flat power-law spectrum is similar to those of high-mass X-ray binaries, 
and the luminosity is within the range of systems with wind-fed accretion.
The presence of a compact object in the system would require a progenitor with an initial mass greater than the initial mass of 
\etacar, i.e. $>$150~\UNITSOLARMASS.  
The evolution of such a massive progenitor would probably result in the creation of a black hole.

\section{Conclusion}

We analyzed datasets of the 10 \SUZAKU\ observations of the super massive star \etacar\ and studied
the variation of the extremely hard X-ray emission above 15~keV through the orbital cycle for the first time.
Our study suggests that the 15$-$25~keV emission originates in the tail of the thermal emission seen below 10~keV,
while the emission above 25~keV is the power-law component observed with \INTEGRAL.
The origin of the power-law component is mysterious.
The $K\alpha$ and $K\beta$ lines of Fe and Ni ions need emission from both CE and NEI plasmas.
The NEI plasma ratio increases toward periastron; this result may suggest an increase of gas density around the WWC apex around periastron.
In the summer of 2014, another X-ray observing campaign for the latest periastron passage is being performed with
multiple X-ray observatories including \NUS, which provides focussed imaging up to 80 keV.
These observations should provide the best measure of the presence of the deeply embedded X-ray component and the power-law component
and their spectral properties (e.g., \NH, \KT, $\Gamma$).
They should help understand the nature of the WWC emission around periastron and the mysterious power-law source.

\acknowledgments

This research has made use of data obtained from the High Energy Astrophysics Science Archive
Research Center (HEASARC), provided by NASA's Goddard Space Flight Center.
This research has made use of NASA's Astrophysics Data System Bibliographic Services.
We appreciate the \SUZAKU\ operations team for optimizing the observations, 
and Masahiro Tsujimoto, Keith Arnaud and Adam Foster for suggestions on the XIS data analysis and the appropriate NEI model.

Facilities: \facility{Suzaku (XIS,HXD)}, \facility{RXTE (PCA)}, \facility{CHANDRA (HETG)}


\end{document}